\documentstyle[12pt,epsfig]{article}
\def\be{\begin{equation}}
\def\ee{\end{equation}}
\def\bea{\begin{eqnarray}}
\def\eea{\end{eqnarray}}
\def\ga{\mathrel{\raise.3ex\hbox{$>$\kern-.75em\lower1ex\hbox{$\sim$}}}}
\def\la{\mathrel{\raise.3ex\hbox{$<$\kern-.75em\lower1ex\hbox{$\sim$}}}}
\def\any{\mathrel{\raise.3ex\hbox{$<$\kern-.75em\lower1ex\hbox{$>$}}}}

\newcommand{\beq}{\begin{equation}}
\newcommand{\eeq}{\end{equation}}
\def\beqa{\begin{eqnarray}}
\def\eeqa{\end{eqnarray}}
\def\p{\partial}

\begin{document}
\begin{titlepage}
\rightline{CERN-TH/2003-140}
\rightline{hep-th/0307201}
\rightline{July 2003}
\begin{center}
 
\vspace{1cm}
 
\large {\bf Quest for Localized 4-D Black Holes in Brane Worlds. II :\\
Removing the bulk singularities.}

\vspace*{10mm}
\normalsize

{\bf  P. Kanti$^{1}$},  {\bf  I. Olasagasti$^{2}$} and 
{\bf K. Tamvakis$^{1,2}$}

\smallskip
\medskip
$^1${\it CERN, Theory Division,\\
CH-1211 Geneva 23, Switzerland}

\vspace*{3mm}                      
                        
$^2${\it Physics Department, University of Ioannina,\\
GR-451 10 Ioannina, Greece}

\smallskip
\end{center}
\vskip0.6in
 
\centerline{\large\bf Abstract}

We analyze further the possibility of obtaining localized black hole solutions
in the framework of Randall-Sundrum-type brane-world models. We consider black
hole line-elements analytic at the horizon, namely, generalizations of the
Painlev\'e and Vaidya metrics, which are taken to have a {\textit{decaying}}
dependence of the horizon on
the extra dimension. These backgrounds have no other singularities apart from
the standard black hole singularity which is localized in the direction of
the fifth dimension. Both line-elements can be sustained by a regular,
shell-like distribution of bulk matter of a non-standard form. Of the two,
the Vaidya line-element is shown to provide the most attractive, natural
choice: despite the scaling of the horizon, the 5D spacetime has the
same topological structure as the one of a RS-Schwarzschild spacetime 
and demands a minimal bulk energy-momentum tensor.

\end{titlepage}

\section{Introduction}
Higher-dimensional models of Gravitation have attracted considerable interest
in the last few years, mostly motivated by
the need  to explain the large difference in magnitude between the Planck
scale $M_P\sim 2\times 10^{18}$ GeV and the Electroweak scale of particle
physics. A number of interesting theoretical ideas have arisen from this
framework \cite{large1,RS1,RS2}. Of particular interest is the
idea of an extra non-compact dimension \cite{RS2} realized in the models
proposed by Randall and Sundrum (RS), where the spacetime corresponds to
regions of AdS space separated by zero-thickness {\textit{$3$-branes}}.
The Standard Model interactions are confined on such a
3-brane, while gravitation, propagating in the five-dimensional
{\textit{bulk}}, is represented on the brane by an ordinary massless graviton
dynamically localized on it \cite{graviton}. The four-dimensional Planck mass
is an effective scale given in terms of the fundamental scale of
five-dimensional gravity and the AdS radius of the five-dimensional spacetime. 
Apart from the study of their cosmological implications, the above
higher-dimensional models have triggered an intense research activity 
on the topic of black holes in the context of scenarios with both
warped~\cite{bh-brane1}\cite{bh-brane2}\cite{bh-nume} and large extra
dimensions~\cite{large}.   

In the standard four-dimensional world, gravitational collapse is described
by the four-dimensional Schwarzschild metric. In a five-dimensional framework,
it would be natural to expect that the (3+1)-dimensional spacetime could still
be described by a Schwarzschild-like metric when matter localized on the brane
undergoes gravitational collapse. For instance, the one-brane RS model
accepts a RS--Schwarzschild black hole solution with a factorized
metric $g_{MN}=Diag(e^{-2\lambda|y|}g_{\mu\nu}^{(S)},\,1)$, with
$g_{\mu\nu}^{(S)}$ the standard four-dimensional Schwarzschild metric. This
solution corresponds to a {\textit{black string}}, infinite in the fifth
dimension~\cite{CHR}, that reduces to a usual Schwarzschild metric on the
brane. Although the five-dimensional Ricci scalar and the square of the Ricci
tensor corresponding to this solution are everywhere finite, the square of the
Riemann tensor diverges at the AdS horizon as $e^{4\lambda |y|}/r^6$. This
singularity renders the above solution physically unsuitable and has led to
the speculation that there exist localized {\textit{black cigar}} solutions
of a finite extension along the fifth dimension arising from a Gregory-Laflamme
type of instability~\cite{GL} near the AdS horizon.

The possibility of obtaining localized black hole solutions in brane models
was investigated in a recent article by two of the present authors \cite{KT}.
There, in the framework of a five-dimensional theory with a warped metric of
the Randall-Sundrum type $g_{MN}=Diag(e^{2A(y)}g_{\mu\nu},\,1)$, a dependence
of $g_{\mu\nu}$ on the extra dimension was introduced. A particular ansatz
corresponding to a 4D Schwarzschild black hole with a ``decaying" horizon
scaled as $r_h(y)=r_h e^{-a y^2}$ was employed. The bulk energy-momentum tensor
sustaining such a behaviour was derived and was shown not to correspond to that
arising from a conventional form of bulk matter (scalar or gauge field). 
For such an exotic, shell-like matter distribution in the bulk, the black hole
singularity is indeed localized near the brane and the spacetime is well
defined near the AdS horizon, in contrast to the behaviour found in
black-string-type solutions.

Nevertheless, the above behaviour was plagued by the appearance of a second,
although localized, singularity at the black hole horizon due to the singular
behaviour of the bulk energy--momentum tensor at this point. An alternative
choice for the five-dimensional metric, which describes a spacetime with a
distinctly different topology from the first one but which also reduces
to the same black hole solution on the brane, was shown to be free of this
singularity. Unfortunately, this choice failed to avoid the singularity at
the AdS horizon. Thus, it was speculated but not demonstrated that the 
additional bulk singularity located at the black-hole horizon is not a
generic feature of every five-dimensional spacetime that induces a black
hole solution on the brane, but that an alternative, well-defined 5D
non-factorized spacetime indeed exists, sustained by an equally well-defined
distribution of bulk matter, possibly of a non-standard form. 

In the present article, we pursue further the above open question. We first
demonstrate that the non-factorized embedding of a non-analytic 4D black-hole
line-element (i.e. with a diverging, at the horizon, $g_{rr}$ metric component),
in a 5D spacetime, results into the manifestation of the black hole horizon
as a true spacetime singularity in the five-dimensional curvature invariant
quantities. We then proceed to introduce alternative metric
ans{\"{a}}tze that describe the same four-dimensional background but which
are analytic at the black hole horizon, namely, the Painlev\'e and
Vaidya spacetimes. A dependence on the fifth dimension of a {\textit{decaying}
form is introduced into the horizon as in the previous study of the
problem~\cite{KT}. Both ans{\"{a}}tze lead to five-dimensional backgrounds
with no other singularity apart from the standard black hole singularity
at $r=0$, which, in addition, turns out to be localized near the brane along
the fifth dimension. These backgrounds are promoted into solutions of 
Einstein's equations of motion if we introduce a bulk energy-momentum tensor
resulting from a shell-like matter distribution localized near the brane and
not attributable to standard forms of bulk matter (scalar, gauge fields etc.).
The Painlev\'e ansatz demands the introduction 
of a complicated bulk energy-momentum tensor, that is well defined 
at the black hole horizon.
 On the other hand, the Vaidya ansatz leads to a minimal model: despite the
non-trivial profile of the black hole horizon along the extra coordinate,
all curvature invariants for this background have the form of the ones of
a factorized RS-Schwarzschild spacetime; the simplicity of the resulting
background is then shown to demand the introduction of only two non-trivial
pressure components in the bulk, a feature often encountered
in brane-world models.


\section{Non-factorized RS-Schwarzschild spacetimes}

In what follows we shall review, in a relatively self-contained way, the
results obtained in the framework of a non-factorized 
Randall-Sundrum-Schwarzschild spacetime, an ansatz introduced in 
Ref. \cite{KT}. In addition, we shall consider generalizations of the
previously studied ans{\"{a}}tze and examine their singularity properties.
The results of this analysis will pave the way for the study of black
hole metrics which are analytic at the horizon and describe localized
black hole configurations.

Looking for black hole solutions that are localized around the
brane, we introduce for the metric a non-factorizable ansatz of the form
\be
ds^2=g_{MN}\,dx^M dx ^N=e^{2A(y)}\hat{g}_{\mu\nu}(x,y)\,dx^{\mu}dx^{\nu}+dy^2\,.
\label{non-fact0}
\ee
Note that a $y$-dependent coefficient of $dy^2$ has been set to unity through
a Weyl rescaling.  The $\hat{g}_{\mu\nu}(x,y)$ metric tensor is
assumed to reduce to the usual Schwarzschild one at the location of the brane
at $y=0$, i.e. $\hat{g}_{\mu\nu}(x,0)=g_{\mu\nu}^{(S)}(x)$,
while the $y$-dependence will provide an additional scaling of the
value of the black hole horizon with respect to the extra dimension, independently
of the one of the warp factor. If this $y$-dependence is appropriately chosen,
the horizon may decay at increasing distance from the brane and eventually
vanish  well before the AdS horizon, thus leading to the desirable localization
of the black hole along the extra dimension and the avoidance of the
singularity at the AdS horizon. 

Our general framework will be the five-dimensional gravitational theory
described by the action
\be
S=- \int d^4x\,dy\,\sqrt{-g}\,\biggl\{-\frac{R}{2 \kappa_5^2}
+ \Lambda_B + \frac{\sigma}{\sqrt{g_{55}}}\,\delta(y) - {\cal L}_B \biggr\}\,,
\label{action1}
\ee
where $\kappa_5^2=8\pi G_5$, with $G_5$ the five-dimensional Newton's constant,
and $\Lambda_B$ stands for a {\textit{bulk cosmological constant}}. In addition,
${\cal L}_B$ represents any additional existing {\textit{bulk matter}}, and the
constant $\sigma$ denotes the positive {\textit{tension}} of the brane located,
as noted above, at $y=0$. 

In Ref. \cite{KT}, we concentrated on the case of a spherically-symmetric
line-element on the brane, and chose the ansatz
\be
ds^2=e^{2 A(y)}\,\biggl\{-U(r,y)\,dt^2 + U^{-1}(r,y)dr^2 + 
r^2(d\theta^2 + \sin^2\theta\,d\varphi^2) \biggl\} \,+ \,dy^2\,.
\label{metric}
\ee
We also assumed that the five-dimensional spacetime is ${\cal{Z}}_2$-symmetric and
therefore invariant under the mirror transformation $y \rightarrow -y$. For
$U=U(r)$ and a vanishing cosmological constant $\Lambda$ on the brane, the
above metric ansatz reduces to the {\textit{black string}} solution~\cite{CHR}
with $A(y)=-\lambda |y|$, where $\lambda^2=\kappa_5^2 |\Lambda_B|/6$, and 
$U(r)=1-\frac{2M}{r}$. For $\Lambda \neq 0$, this solution can be easily
generalized to the RS--AdS/dS--Schwarzschild one \cite{RS-AdS}.
Both types of five-dimensional solutions have finite $R$ and $R_{MN} R^{MN}$,
while the square of the Riemann tensor is given by 
\be
R_{MNRS}R^{MNRS}\propto \frac{48 M^2 }{r^6}\,e^{-4A(y)}\,.
\label{riemann}
\ee
The above reveals the existence of a string-like black hole singularity
infinitely extending along the extra dimension as well as an additional
singularity at the point where the warp factor vanishes, either at the AdS 
horizon for a flat brane, or at a finite coordinate distance for a de Sitter
brane. 

The introduction of $y$-dependence in the metric function $U$ and, therefore,
the resulting scaling of the horizon or, equivalently, the ``mass'' of the black
hole, with the distance from the brane, could provide a resolution to both
of the above problems. If $M(y)$ decreases faster than the function
$e^{-2 A(y)}$ increases, then the singular term in Eq. (\ref{riemann}) vanishes
before the spatial limit of the extra dimension is reached. Unfortunately,
as it was shown in \cite{KT}, an empty bulk or a bulk with a single scalar
or gauge field cannot support such a $y$-dependence. We then followed an
alternative approach by demanding an exponentially decaying black hole horizon,
i.e. $U(r,y)=1-w(y)/r$, with $w(y)=2 M e^{-a y^2}$ and determining the
components of the unknown
bulk energy--momentum tensor through Einstein's equations. This particular
ansatz was found to demand the introduction of three independent non-trivial 
components of the bulk energy-momentum tensor, namely $T^t_t$, $T^r_r$ and
$T_{r5}$. All three were shown to correspond to a localized shell-like
distribution of exotic matter around the brane and to vanish at both $r$
and $y$ infinity.

Although the singular term in Eq. (\ref{riemann}) decayed exponentially with
the distance from the brane, thus, localizing the black hole singularity and
removing the one from the boundary of the extra dimension, a new problem arose.
The harmless, in 4 dimensions, horizon non-analyticity at $r=w(y)$ was transformed
into a true singularity of the five-dimensional spacetime. All components of the
bulk energy-momentum tensor diverged at the location of the horizon with the
same singularity arising in all curvature invariant quantities through the
appearance of additional terms singular at $r=w(y)$. For instance, the scalar
curvature took the form $R=-20 A'^2 - 8 A'' -w'^2/2(r-w)^2$. 
Although all singular terms appearing in the invariant quantities also
decay exponentially away from the brane, the question of whether the appearance
of a second singularity is unavoidable in trying to localize a bulk string-like
black-hole singularity remained open.

We will now show that the embedding in a higher-dimensional spacetime of
a brane line-element that is characterized by the presence of a horizon and of
{\it mixed} coordinate dependence, i.e. dependence on both brane and bulk
coordinates,quite generally results in the manifestation of brane horizons as true,
bulk singularities of the higher-dimensional theory. This result turns out to
be independent of the dimensionality of the problem and can be demonstrated for a more general
 four-dimensional metric ansatz than the one initially chosen. We may therefore consider
the following Schwarzschild-like line-element
\be
d \hat s^2_p=-U^2(r,y)\,dt^2 + \frac{dr^2}{V^2(r,y)} + r^2\,d\Omega^2_{p-2}\,,
\label{metric-g}
\ee
describing a $(p+1)$-dimensional brane,  embedded in a $D$-dimensional spacetime
of the form 
\be
d s^2=g_{MN}\,dx^M dx^N = e^{2 A(y)}\,\hat g_{\mu\nu}(x,y)\,dx^\mu dx^\nu
+ \gamma_{mn}(y)\,dy^m dy^n\,.
\label{metricD}
\ee
The existence of true spacetime singularities will be reflected in the expressions
of the $D$-dimensional curvature invariant quantities. For our purposes, the
expression of the scalar curvature will be adequate. In the Appendix, we give
the non-vanishing Christoffel symbols and the components of the Ricci tensor
that lead to the evaluation of this quantity. 

In the restricted case where $V=U$,  the determinant $\sqrt{-\hat{g}}$ is
$y$-independent, and, therefore, all terms in the expression of the scalar
curvature in the Appendix involving derivatives of $\ln \sqrt{-\hat{g}}$
with respect to the extra coordinates will vanish. Thus, the singular terms
in the expression of $R$ are 
\be
\frac{1}{4}\,\hat{g}^{\mu\nu}\hat{g}^{\rho\lambda}
\hat{g}_{\nu\lambda,m}\hat{g}_{\mu\rho}^{\ \ \,,m}=
-\frac{2}{U^2}\,\gamma^{mn}\partial_mU\,\partial_nU\,.
\ee
Since the $(p+1)$-dimensional metric describes a Schwarzschild-like black hole
on the brane, $U(r,y)$ must vanish at the horizon. Then, the
above quantity diverges transforming the $(p+1)$-dimensional coordinate
singularity to a true singularity of the $D$-dimensional spacetime. A special
case of the above is the ansatz (\ref{metric}) used in \cite{KT} where the
presence of a horizon on the 3-brane led indeed to the appearance of a second
bulk singularity.

Let us go back to the general case of different $U$ and $V$ metric functions
and let us put together all terms from Eq. (\ref{scalarR}) involving solely
derivatives of the metric tensor $\hat g_{\mu\nu}$ and its determinant,
which is now obviously $y$-dependent. Then, we end up with
\bea
&~& \hspace*{-0.9cm} -2 \tilde D^m \tilde D_m \ln\sqrt{-\hat g} -
\partial^m \ln\sqrt{-\hat g}\,\partial_m \ln \sqrt{-\hat g} +
\frac{1}{4}\,\hat g^{\mu\nu} \hat g^{\rho\lambda}\,\hat g_{\nu\lambda,m}\,
\hat g_{\mu\rho}^{\ \ \,,m}
+ \frac{1}{2}\,\hat g_{\mu\nu,m}\,\hat g^{\mu\nu,m}=\nonumber \\[2mm]
&~& \hspace*{1.7cm} =-2\gamma^{mn}\,\biggl(\frac{\partial_m\partial_n U}{U}-
\frac{\partial_m\partial_n V}{V} +2 \frac{\partial_m V}{V}\,
\frac{\partial_n V}{V}-\frac{\partial_m U}{U}\,
\frac{\partial_n V}{V}\Biggr)\,.
\label{add2}
\eea
Another term that might also diverge at the location of the horizon is
proportional to:
\be
\partial^m A\,\partial_m \ln \sqrt{-\hat g} =\partial^m A\,
\biggl(\frac{\partial_m U}{U}-\frac{\partial_m V}{V}\biggr)\,.
\label{add3}
\ee
If we want the metric functions to be chosen independently of the warp
factor function $A(y)$, then, the singular combination inside brackets in
the last expression must vanish. In the case, where only $U$ vanishes, or
$U$ and $V$ vanish at different values of $r$, the singularity is unavoidable.
In the case, where both $U$ and $V$ vanish at the same value of $r$, the
only solution that leads to the vanishing of this combination term is:
$U(r,y)=g(r)\,V(r,y)$. But then, Eq. (\ref{add2}) is always plagued
by a singularity. If on the other hand, we allow
for the metric functions and the warp factor to be related, then, it is
the combination of Eqs. (\ref{add2}) and (\ref{add3}) that must vanish
instead. In principle, one cannot exclude the possibility that a metric
ansatz, that could render the scalar curvature finite, may be found.
Nevertheless one has to ensure that all curvature quantities become finite
at the same time. Despite our attempts, we have not been able to find
such a metric ansatz.

Clearly, the key to the removal of the additional bulk singularities is the
use of a brane line-element that describes
a black hole solution but that is analytic at the horizon.
In four dimensions, one can use
different line-elements
related by coordinate transformations in order to describe the
same spacetime. 
However, the embedding of these line-elements in an extra
spacetime automatically breaks their equivalence. The resulting
higher-dimensional line-elements no longer describe the same spacetime, a
fact which is reflected in the different sets of curvature invariant quantities
determined for each line-element.  In \cite{KT}, a Kruskal-Szekeres brane
line-element, that also describes a black hole solution but without possessing
a horizon, was used instead of the Schwarzschild one. It was shown that
the corresponding five-dimensional spacetime was free of the presence
of the additional singularity. Nevertheless, this particular choice failed
to either eliminate the singularity at the AdS horizon or to localize the
black hole singularity. In the next two sections, we will study two
alternative, analytic, five-dimensional line-elements, with different
topological structure but both describing a black hole on the brane, that
will be shown to achieve all of the above goals.

%
%

\section{The Non-factorized RS-Painlev{\'{e}} spacetime}

Let us start by considering a five-dimensional non-factorized metric ansatz 
that contains a Painlev{\'{e}}-type brane line-element \cite{PAIN}. The
four-dimensional Painlev{\'{e}} line-element has been more recently
employed in the study of black hole quantum mechanics \cite{KW}. A
non-factorized 5D version of it can be written as
\be
ds^2=e^{2A(y)}\biggl[-\Bigl(1-\frac{2m(y)}{r}\Bigr)\,dv^2 -
2\,\sqrt{\frac{2 m(y)}{r}}\,dv\,dr+ dr^2 +
r^2 (d\theta^2 + \sin^2\theta\,d\varphi^2)\biggr] + dy^2\,.
\label{pain-m}
\ee
At the location of the brane, at $y=0$, we demand the mass function to reduce
to the usual ADM mass parameter of the black hole, i.e. $m(0)=M$. Then, the
induced metric on the brane takes the form of the Schwarzschild solution under
the coordinate transformation 
\be
v \,\,\longrightarrow \,\,t=v + 4M\,\biggl(\sqrt{\frac{r}{2M}} - 
{\rm Arctanh}\sqrt{\frac{r}{2M}}\,\,\biggr)\,.
\label{pain-c}
\ee
The advantage  of the line-element (\ref{pain-m}) compared to the one given
in Eq. (\ref{metric}) is the fact, that although both describe the same
spacetime, the former is analytic at the horizon, i.e. no divergences appear
in the metric components at $r=m(y)$.

 Hopefully, that
will lead to the avoidance of the additional bulk singularity that plagued
the Schwarzschild choice. On the
other hand, the $y$-dependence of the mass function, or equivalently
of the horizon, will be used in the same way in order to localize the
black-hole singularity close to the brane and resolve the singularity at
the AdS horizon. 

From the five-dimensional perspective, the metric ansatz (\ref{pain-m}) must
satisfy Einstein's equations that are found to have the form
\bea
6 A'^2 + 3 A'' + \frac{2 A' m'}{r} + \frac{m''}{2 r} - \frac{m'^2}{8 r m} 
&=&\kappa_5^2 \,\Bigl[-\Lambda_B + S^v_{v}\,\delta(y) + \hat T^v_{v}] 
\label{pain-1}\\[4mm]
6 A'^2 + 3 A'' -\frac{2 A' m'}{r}  - \frac{m''}{2 r} -\frac{m'^2}{8 r m} 
&=&\kappa_5^2 \,\Bigl[-\Lambda_B + S^r_{r}\,\delta(y)  + \hat T^r_{r}\Bigr] \\[4mm]
6 A'^2 + 3 A'' -\frac{m'^2}{8 r m}
&=&\kappa_5^2 \,\Bigl[-\Lambda_B + S^\theta_{\theta}\,\delta(y)  + \hat T^\theta_\theta\Bigr] \\[4mm]
6 A'^2 + \frac{m'^2}{8 r m} &=&\kappa_5^2 \,(-\Lambda_B + \hat T^y_{y}) \\[4mm]
-\sqrt{\frac{2 m}{r}}\,\biggl(\frac{A' m'}{m} + \frac{m''}{4 m} -
\frac{m'^2}{8 m^2}\biggr)&=&
\kappa_5^2 \,\Bigl[S^v_r\,\delta(y)+\hat T^v_{r}\Bigl] \\[4mm]
-\frac{m'}{4r^2}\,\sqrt{\frac{2 m}{r}}\,\Biggl(1+\frac{3r}{2m}\biggr)
&=&\kappa_5^2  \,\hat T_{vy} \\[4mm]
\frac{m'}{4r^2}&=& \kappa_5^2  \,\hat T_{ry}\,, \label{pain-7}
\eea
where $S_{\mu \nu}$ is the energy-momentum tensor localized on the brane,
the simplest case being that of a brane with a constant tension, namely, 
$S_{\mu}^{\nu}=\sigma\,h_{\mu}^{\nu}$. We will see below that for a general
mass function $m(y)$ this may be non-trivial, that is, the brane may not 
be a vacuum wall. In the above, we have also assumed that the {\it total}
energy-momentum tensor in the bulk, $T_{MN}$, can be written as the sum of
two contributions, one coming from the bulk cosmological constant and one
from the presence of a bulk matter distribution related to ${\cal L_B}$,
that is
\be
T_{MN}= - g_{MN} \,\Lambda_B + \hat T_{MN}\,.
\ee
For the remaining of this paper, we will be assuming that a negative bulk
cosmological constant gives rise to the usual Randall-Sundrum form of the
warp factor $e^{A(y)}=e^{-\lambda |y|}$. Then, the terms on the {\it lhs}
of the diagonal components of Einstein's equations that are proportional to
$A'^2$ and $A''$ are cancelled by the presence of $\Lambda_B$ on the
{\it rhs}. All remaining terms containing derivatives of the metric function
$m(y)$ are attributed to the presence of the bulk energy-momentum tensor
$\hat T_{MN}$. In other words, the non-trivial profile of the horizon of the
black-hole along the extra dimension demands the introduction of a bulk
matter distribution. For this particular metric ansatz, it is necessary to
introduce non-trivial bulk energy-density and diagonal pressure components
as well as three off-diagonal ones.  Since we are reading the 
energy-momentum tensor from the Einstein equations, the conservation
equation  $D_M T^M_N=0$ is trivially satisfied for an arbitrary function $m(y)$.

Having introduced a brane line-element with no apparent horizon, we expect
that no additional singularity, apart from the black-hole one, will plague
the five-dimensional spacetime. By looking at the components of the 
energy-momentum tensor, we can immediately see that they do not diverge
at any other point apart from the one at $r=0$. Nevertheless, we should
check the expressions of the curvature invariant quantities which are directly 
related to the singularity structure of spacetime.
 The scalar
curvature is
\be
R = -20 \lambda^2 + \frac{m'^2}{4 r m}\,, 
\ee
while the square of the Riemann tensor has the form
\bea
\hspace*{-0.5cm} R_{MNRS} R^{MNRS} 
&=& 40 \lambda^4 + \frac{48\,m^2\,e^{-4 A}}{r^6} + 
\frac{A' m'}{r m}\,\Bigl(\frac{2 m'^2}{m} - 7A' m' - 4 m'' \Bigr)
\nonumber \\[2mm] \hspace*{-0.5cm} &+&
\frac{m'^2 e^{-2 A}}{2 r^4}\,\biggl(13-\frac{9 r}{2 m}\biggl) 
+\,\frac{m'^2}{rm^2}\,\biggl(m''- \frac{m'^2}{4 m}
+\frac{11 m'^2}{16 r}\biggr) - \frac{m''^2}{r m} \,.
\eea
From the expressions of the above quantities \footnote{The square of the Ricci
tensor, $R_{MN} R^{MN}$, has a form similar to the one of the Riemann tensor
but lacks the $1/r^6$ term.}, we may see that {\textit{the only singularity
that makes its appearance in the scalar gravitational quantities is the black
hole singularity at $r=0$}}. The mass function $m(y)$ is assumed to be everywhere
well-defined, to scale faster than $e^{2 \lambda |y|}$ and to reduce to zero away
from the brane. In this way, all singular terms, both at $r=0$ and at the AdS
horizon $|y| \rightarrow \infty$, vanish at a moderate distance from the brane
thus restoring a regular five-dimensional spacetime.

An ansatz that manages to accommodate all the above demands and
reduces to the usual mass parameter $M$ on the brane is:
$m(y)=M\,e^{-a |y|^{n}}$ where $n$ is an integer number and $a$ a positive
constant. Depending on the value of $a$, the localization of the singular terms,
appearing in the curvature invariant quantities, takes place at different
distances from the brane but always before the AdS horizon is reached; for
$a \gg 2 \lambda^{2-n}$, this localization takes place at a distance smaller
than the AdS radius, $\lambda^{-1}$. The above ansatz
introduces no discontinuity in the mass function as long as $n>1$ since in
this case $m'(0)=0$. Discontinuities of the derivatives of the metric tensor
are directly related to a localized matter tensor on the brane through the
expression
\beq
[K]_{\mu \nu}-h_{\mu \nu} Tr[K] =-\kappa^2_5 S_{\mu \nu}\,,
\eeq
$K_{\mu\nu}$ being the extrinsic curvature and $h_{\mu\nu}$ the induced
metric on the brane. The brackets denote the discontinuity across
the brane. Taking into account the ${\cal{Z}}_2$ symmetry, we get
\beq
[K]_{\mu \nu}=\left[2 A' h_{\mu \nu}
+ \delta^v_\mu \delta^v_\nu\; {2m'\over r}
-{1\over 2}(\delta^r_\mu \delta^v_\nu+\delta^v_\mu \delta^r_\nu)
\; \sqrt{2m\over r}\; {m'\over m}\right]_{y=0^+}\,.
\eeq
The first term in the above expression is the standard RS tension term
proportional to $h_{\mu\nu}$ that causes the discontinuity in the second
derivative of the warp factor. Nevertheless, for $m'(y=0)\neq 0$, additional
terms could also be present. This is indeed the case for $n=1$: the
discontinuity that arises can be accommodated only in the presence of additional
matter strictly localized on the brane. Although this case cannot be excluded,
we consider it as a rather complicated and a physically unrealistic one.

Finally we need to check the profile of the bulk energy-momentum tensor
components.
By mere inspection of the field equations (\ref{pain-1})-(\ref{pain-7}), we
may see that, for fixed $y$, the components of the energy-momentum tensor
scale as $1/r^k$, where $k$ a positive number, therefore all of them go to
zero at large distance from the horizon of the black hole. On the other hand,   
a physically sensible distribution of bulk matter demands also its localization
close to the brane: a bulk matter distribution diverging at $y$-infinity
would be in contradiction with the concept of localization of gravity around
the brane. Figure 1 depicts the profile of the five independent ``extra''
components along the $y$-axis for three different values of $n$. We choose,
for simplicity, $r=r_h=1$ and $\lambda=1$, in units of Planck mass. 
The case $n=1$ is 
qualitatively different from the one in which $n>1$ also when it comes to
the distribution of bulk matter: all of the components are peaked at the
location of the brane and rapidly reduce to zero as $y$ increases. On the other
hand, for $n>1$, all components reach a maximum, in absolute value, at a
location off the brane before reducing to zero. Small differences appear
between the cases $n=2$ and $n>2$: in the latter case, all components
vanish at the position of the brane while, in the former one, some of the
bulk components, including the energy-density, adopt non-zero values at the
position of the brane without, however, causing any discontinuity in the
Einstein tensor. 

In all cases considered, the bulk matter is localized either on or off the
brane while vanishing at infinity, thus creating a shell-type distribution
of bulk matter. For $n>1$, the energy-density $\rho=-\hat T^v_v$ and radial
pressure $p_r=\hat T^r_r$ satisfy a stiff equation of state, i.e. 
$\rho \simeq p_r$, however, they are not the dominant components as
the off-diagonal component $\hat T^v_r$ prevails both at small and large
distance from the brane. Almost all of the components, including the
energy-density, change sign at some point in the bulk, and thus one can
easily show that both the weak and strong energy conditions are violated
at specific regimes in the bulk. This behaviour forces us to interpret
them as arising from a non-ordinary distribution of matter. 
The importance of the non-factorized RS-Painlev\'e metric ansatz lies on the
fact that it constitutes the first example of a spacetime with a $y$-dependent
horizon that is well-defined for every value of the radial coordinate $r>0$.
More specifically, no divergences appear either in the components of $T_{MN}$,
or in the curvature invariant quantities, at the location of the
four-dimensional horizon of the equivalent Schwarzschild solution. The
above example, as well as the one of the generalized Vaidya spacetime to
be studied shortly, clearly confirm our argument that analytic black hole
line-elements do not generate additional bulk singularities. 
%

%
\begin{center}
\begin{figure}[t]
\begin{tabular}{c}
\hspace*{1cm} \epsfig{file=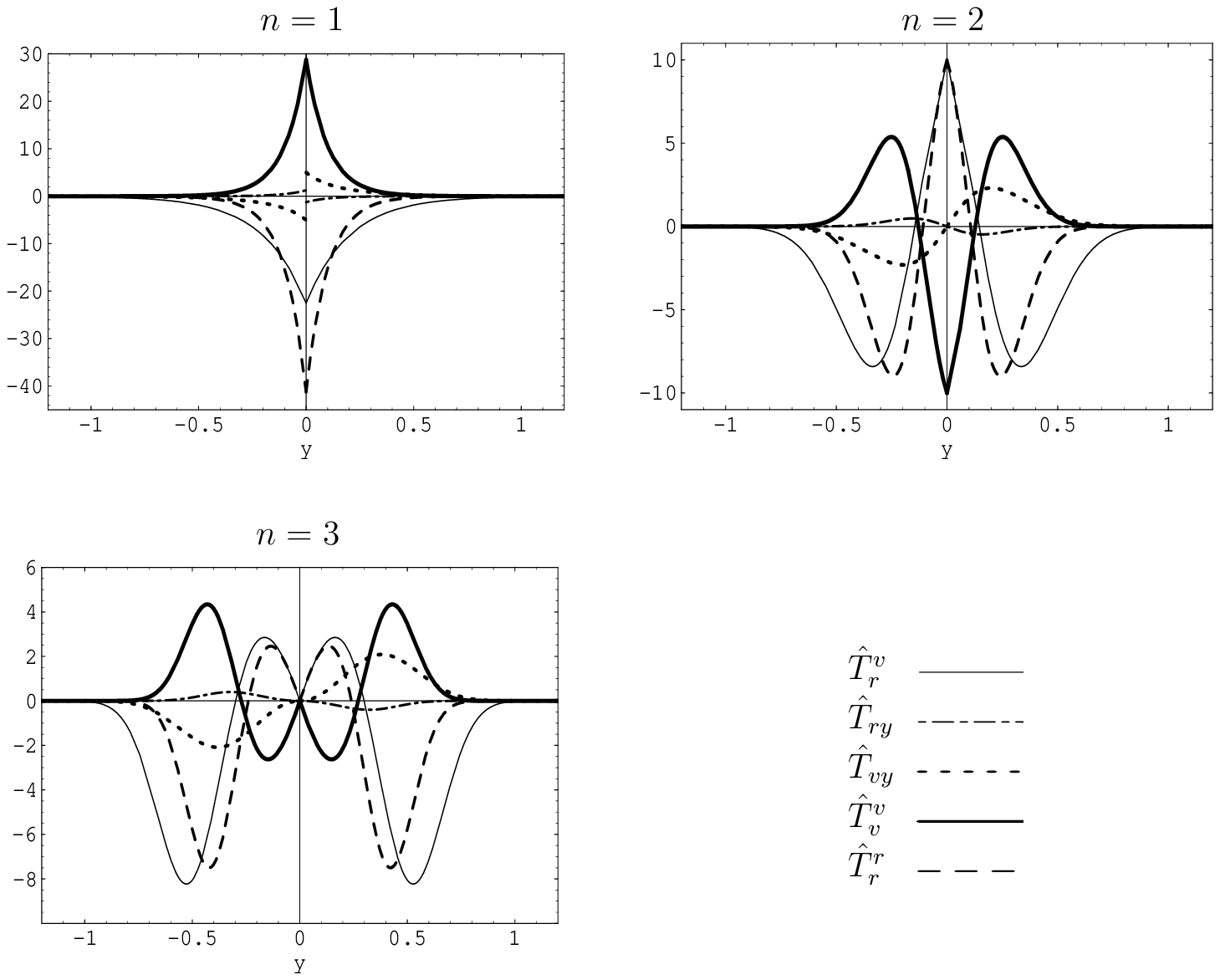, height=11cm}
\end{tabular}\\[3mm]
{{\bf Figure 1:}  {\it The plot depicts the profile of the five independent
components of the bulk energy--momentum tensor along the bulk ($y$) coordinate,
for $\lambda=1$, $a=10$ for $n=1$ and $a=20$ for $n=2,3$  and for fixed $r=r_h=1$. 
All components create a shell-type distribution of matter in the bulk and then
vanish at moderate distances from the brane.}}
\end{figure}
\end{center}

%
%

\vspace*{-10mm}
\section{The Non-factorized RS-Vaidya spacetime}

We will now consider an alternative metric ansatz, that describes a Vaidya-type 
space\-time on the brane, of the form \cite{vaidya}\cite{PI}
\be
ds^2=e^{2A(y)}\biggl[-\Bigl(1-\frac{2m(v,y)}{r}\Bigr)\,dv^2+ 
2 \epsilon dv dr+
r^2 (d\theta^2 + \sin^2\theta\,d\varphi^2)\biggr] + dy^2\,,
\label{vaidya}
\ee
where $\epsilon= \pm 1$. The $y$-dependence of the mass function $m$ is assumed
to satisfy the same criteria as in the previous cases: to become trivial at
the location of the brane and to cause a rapid decrease of $m$ away from it.  
For a $v$-independent mass function, the above line-element takes the form
of the Schwarzschild solution on the brane under the coordinate transformation 
\be
v \,\,\longrightarrow \,\,t=v - r - 2M\,\log(r-2M)\,.
\label{vaidya-c}
\ee
Similarly to the line-element presented in Eq. (\ref{pain-m}), the above metric
tensor has no apparent singularity at the horizon. Therefore, we expect
that no additional bulk singularities will appear. In addition, the simpler form
of this particular ansatz is quite likely to lead to a simpler form for the bulk
energy-momentum tensor. 

In accordance to the usual Vaidya line-element in four dimensions, we have kept
the $v$-dependence of the mass function. For $\epsilon=+1$, the null coordinate
$v$ represents the Eddington advanced time and the corresponding
line-element may be used to describe the collapse of a spherical shell of matter
on the brane leading to the increase of its mass with time, $\partial_v m>0$. On
the other hand, the choice $\epsilon=-1$ represents the Eddington retarded time
which may be used for the study of an expanding shell of matter. Here, we
will be  interested in the former case, with the $v$-dependent Vaidya
metric describing the spacetime on the brane {\it during} the collapse, and
the $v$-independent one, or its equivalent Schwarzschild metric, describing
the spacetime {\it before} or {\it after} the collapse. In what follows,
unless otherwise stated, the mass function will be assumed to be both
$v$ and $y$-dependent.

Let us first turn to the non-vanishing components of the five-dimensional
Einstein's equations. We find that, for the metric ansatz (\ref{vaidya}), the
diagonal components simply reduce to the usual ones for an AdS spacetime,
namely
\bea
 6 A'^2 + 3 A''&=&\kappa_5^2 \,\Bigl[-\Lambda_B + S^v_{v}\,\delta(y) +
\hat T^v_{v}] \\[4mm]
6 A'^2 + 3 A''&=&\kappa_5^2 \,\Bigl[-\Lambda_B + S^r_{r}\,\delta(y) +
\hat T^r_{r}] \\[4mm]
6 A'^2 + 3 A''&=&\kappa_5^2 \,\Bigl[-\Lambda_B + 
S^\theta_{\theta}\,\delta(y) + T^\theta_\theta \Bigr] \\[4mm]
6 A'^2 &=&\kappa_5^2  \,\Bigl[-\Lambda_B + \hat T^y_{y}]
\eea
where $S_{\mu \nu}$ is again the stress-energy tensor localized on the wall.
If we, therefore, assume that, as before, the bulk cosmological constant gives
rise to the usual Randall-Sundrum form of the warp factor, the bulk energy-momentum
tensor does not need to have non-trivial energy-density and diagonal pressure
components. Nevertheless, we must have two independent non-trivial, off-diagonal
pressure components, defined by the equations
\bea
-\epsilon \biggl(\frac{4 A' \partial_y m}{r} + \frac{\partial^2_y m}{r}\biggr)
+\frac{2 (\partial_v m)\,e^{-2 A}}{r^2} &=& \kappa_5^2  \,\hat T^r_{v}\,, 
\label{Trv} \\[3mm]
e^{-2 A}\,\frac{\partial_y m}{r^2} &=& \kappa_5^2 \,\hat T^r_y\,.
\eea
Once again the equation for the conservation of the energy-momentum tensor,
$D_M T^M_N=0$, is trivially satisfied  by the above components for an
arbitrary function $m(v,y)$. The presence of the bulk components is
necessary to support the non-trivial form of the mass function $m(v,y)$.
Note that even for $\partial_y m=0$, in which case we recover the usual 4D
Vaidya metric for each slice $y=const.$, the necessity of having a {\it bulk
off-diagonal pressure component}, $\hat T^r_{v}$, still arises. We may
therefore conclude that whenever the 4D metric has a structure that demands
a non-trivial energy-momentum tensor, as for example in the case of the
dynamically evolving Vaidya line-element, its consistent embedding in a
higher-dimensional spacetime demands the introduction of a corresponding
bulk energy-momentum tensor. Leaving the
problem of the localization of the black-hole singularity aside, that means
that the process of the production of black holes or, in general, of the
collapse of matter on the brane cannot be a purely four-dimensional process 
that can take place in a empty bulk.

The finiteness for $r > 0$ characterizes not only the components of the bulk
energy-momentum tensor but also the curvature invariants derived for the
line-element (\ref{vaidya}), which are found to be
\bea
R &=& -20 A'^2 - 8 A'' \,, \nonumber \\[4mm]
R_{MN} R^{MN} &=& 4\,(20 A'^4 +16 A'^2 A'' + 5 A''^2) \,, \nonumber \\[2mm]
R_{MNRS} R^{MNRS} &=& 8\,\biggl(5 A'^4 +4 A'^2 A'' + 2 A''^2 +
6\frac{ m^2(v,y)}{r^6}\,e^{-4 A(y)}\biggr)\,,
\eea
for an arbitrary warp factor function $A(y)$. Similarly to the case studied
in the previous section, the above scalar quantities do not have
any new singularities in the bulk. 
Formally they are identical to the ones derived for
a RS-Schwarzschild line-element (although in that case $m(v,y)=m=constant$), 
however, in this case, the scaling of the
horizon with the distance from the brane can lead to the localization of
the black-hole singularity. 

In the following, let us consider $A(y)=-\lambda|y|$ and the simple ansatz
$m(v,y)=m(v)\,e^{-a |y|^{n}}$, where $n$ is an integer number and $a$ a 
positive constant. For large enough values of $a$, the horizon reduces to
zero much faster than the square of the warp factor so that the last, singular
term in the expression of the square of the Riemann tensor goes to zero
as $y\rightarrow \infty$ unlike the black string solution (again, for
$a \gg 2 \lambda^{2-n}$, the localization takes place at a distance smaller
than the AdS radius $\lambda^{-1}$). As in the case
of the Painlev{\'{e}} metric, the simplest choice would be to consider a
mass function continuous in $y$. In the most general case of
discontinuous $m$, and assuming ${\cal{Z}}_2$ symmetry, we have for the
extrinsic curvature
\beq
K_{\mu \nu}=-\Gamma^y_{\mu \nu}=\left. A' h_{\mu \nu}
+\delta^v_\mu \delta^v_\nu \,{\p_y m(v,y)\over r } \right|_{y=0}\,,
\eeq
and for the localized matter tensor
\beq
\kappa^2_5 S_{\mu \nu}=\left[ 6 A' h_{\mu \nu}
-2\delta^v_\mu \delta^v_\nu \,{\p_y m(v,y)\over r } \right]_{y=0^+}\,.
\eeq
For the RS decaying exponential warp factor, we get the usual tension term plus
an additional term for the $S_{vv}$ component provided
$\partial_y m(v,y=0) \neq 0$. For $n>1$, this quantity is zero as before, and
the brane is characterized only by a constant tension. For $n=1$, however, the
mass function is discontinuous on the brane and the field equations demand the
introduction of a null fluid on the brane characterized by a single 
non-vanishing component $S_{vv}$. We consider this case as a non-realistic
one, however, we include it in the following discussion as a choice consistent
with the field equations. 

Both bulk components scale as $1/r$ at large distances from the horizon, taken
to be $r_h=2 m=1$ at a fixed time $v=const$, and thus vanish asymptotically.
In order to study the profile of the components of the bulk energy-momentum
tensor along the extra dimension, we may set $r=r_h$ and $\lambda=1$, again 
in units of Planck mass. Since
we study the collapse of brane matter, we further assume that $\epsilon=+1$
and $\partial_v m>0$. We have denoted as $\hat T^{r (s)}_v$ the ``static part''
of $\hat T^r_v$, i.e. the first two terms in Eq. (\ref{Trv}), that represent
the value of this component in the two static limits, either {\it before} or
{\it after} the collapse of the brane shell. On the other hand, $\hat T^{r (a)}_v$
stands for the total expression containing also the non-static term
proportional to $\partial_v m$, and gives the value of the component
{\it during} the collapsing phase. The profile of the off-diagonal components
of the energy-momentum tensor, $\hat T^r_v$ and $\hat T^r_y$, along the 
$y$-axis, is depicted in Fig. 2. As desired, both components, either
static or non-static, are ``localized'' around the brane and vanish
exponentially fast as $y \rightarrow \infty$. 

\begin{figure}[t]
\begin{tabular}{c c}
\hspace*{1.3cm} {\small $n=1$} & \hspace*{0.6cm} {\small $n=2$} \\[1mm]
\hspace*{1cm}\includegraphics[width=2.5in,height=2in,bb=88 4 376 182]{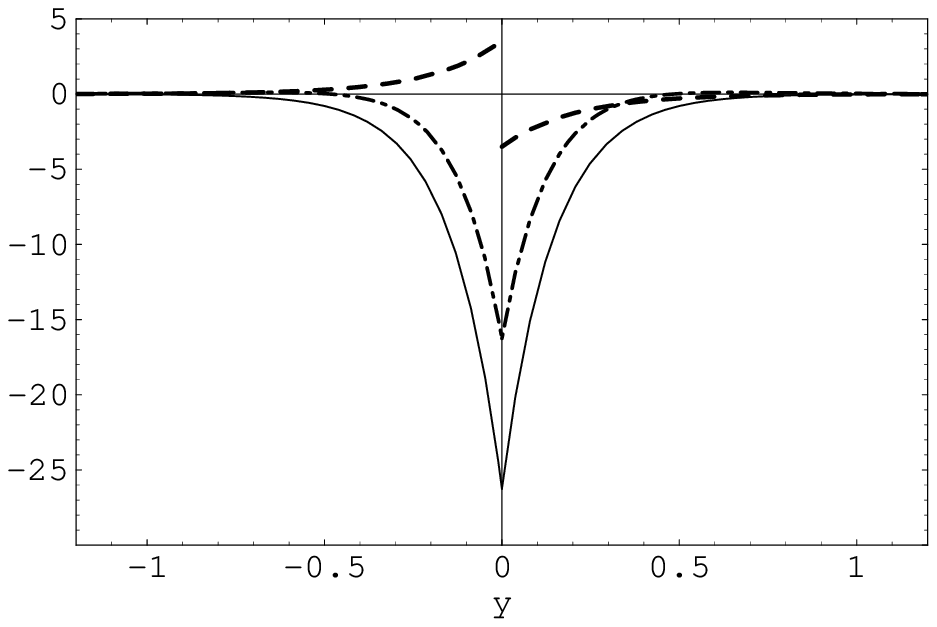} &
\hspace*{0.3cm}\includegraphics[width=2.5in,height=2in,bb=88 4 376 182]{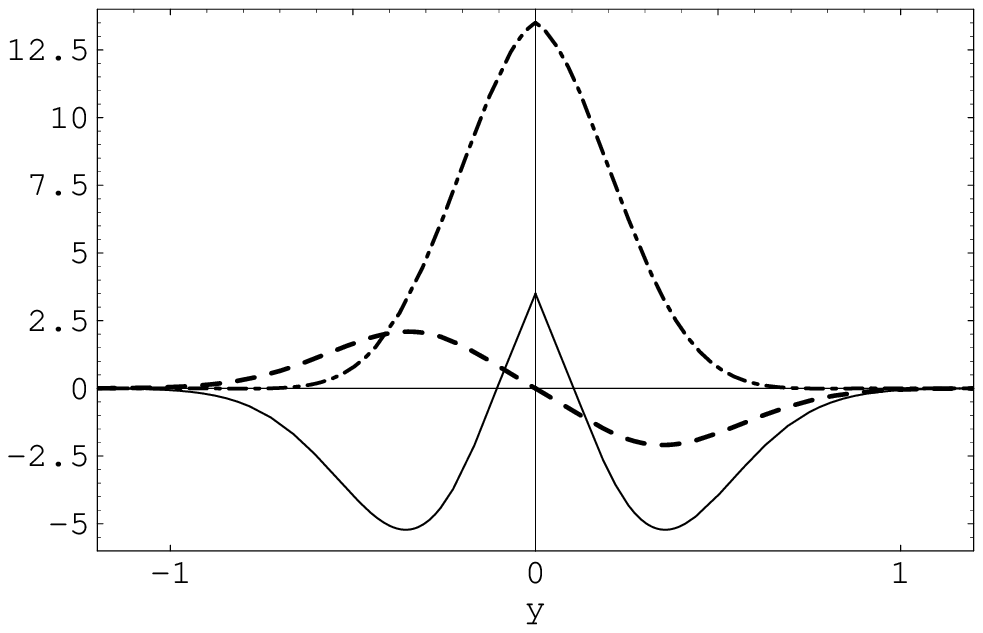}
\end{tabular}
\vspace*{2mm}
\centerline{\begin{tabular}{c}
\hspace*{0.1cm} {\small $n=3$}\\
\includegraphics[width=2.5in,height=2in,bb=88 4 376 182]{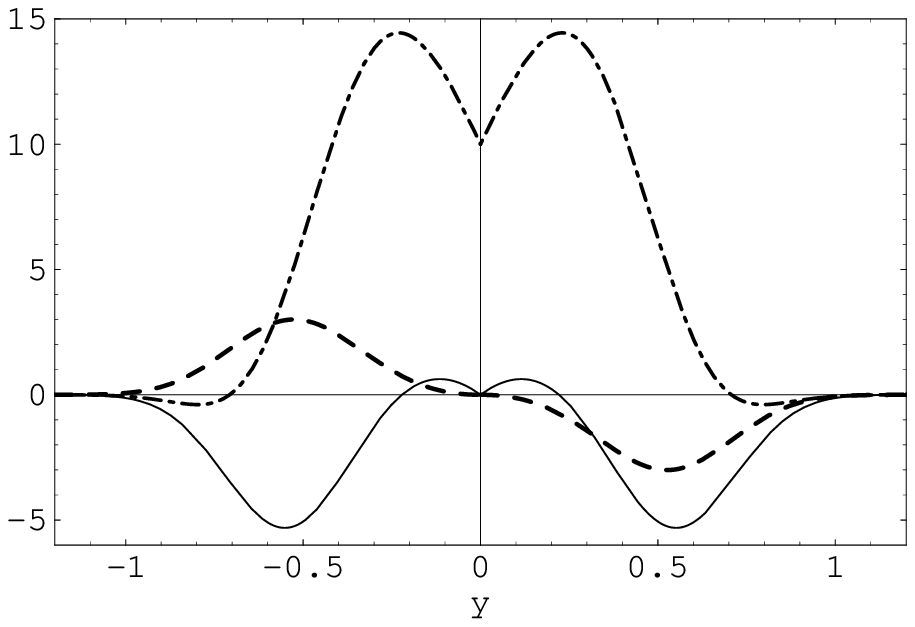} 
\end{tabular}}
{{\bf Figure 2:}  {\it The plot depicts the profile of $\hat T^{r(s)}_v$
(continuous), $\hat T^{r(a)}_v$ (dot-dashed) and $\hat T^r_y$ (dashed)
components of the bulk energy-momentum tensor along the bulk ($y$) coordinate,
for $\lambda=1$, $a=7$  and for fixed $r=r_h=1$. For the non-static
component $\hat T^{r(a)}_v$  we have used $\partial_v m(v)=5$. All components
create a shell-type distribution of matter in the bulk and then vanish at
moderate distances from the brane.}}
\end{figure}

We can see, once again, that the bulk matter distribution is qualitatively different
depending on the value of $n$. For $n=1$, we can see that some bulk components
have a discontinuity at the brane position while, for $n>1$, all of them
are continuous across the brane. Focusing on the more realistic case of $n>1$,
we note that before or after the collapse, when the line-element is taken to
be static, the $\hat T^r_v$ off-diagonal pressure component reaches a negative
minimum value at some distance off the brane before going to zero at larger
distances. As the collapse begins and $\partial_v m$ starts to increase, the
$\hat T^r_v$ component increases too, reaching a maximum positive value off
the brane. As the collapse reaches its end, the $\hat T^r_v$ component falls
back into its static, negative value $\hat T^{r (s)}_v$. We may thus conclude
that while a bulk, negative $\hat T^r_v$ pressure component is necessary to
support a $y$-decaying horizon, a symmetric positive pressure barrier is
necessary in the bulk during dynamical collapse on the brane, when we have a
non-trivial in-flow of energy. The remaining off-diagonal
component $\hat T^r_y$ is not affected by the process of collapse on the brane.

The above bulk energy-momentum tensor has zero energy-density and diagonal
pressure components, so its interpretation by means of an ordinary matter
distribution (a single scalar or gauge field) is not possible. Specific
attempts towards this direction have proved to be unsuccessful similarly to
the analysis performed in \cite{KT}. However, the demand for the existence
of only pressure bulk components, that are not accompanied by an energy-density,
is not unusual in brane models. Such components have been introduced in
single \cite{KKOP} or two brane \cite{KOP} models in order to modify the
topology of the extra dimension. In all those examples, the bulk pressure
component was merely a reflection of what existed or was taking place on
the brane. The present situation is very similar to the aforementioned one
as the introduced pressure components keep the topological structure of a
brane black hole embedded in a higher-dimensional spacetime close to the brane.
In \cite{KKOP}, it was shown that the existence of pressure components
without a corresponding energy-density in the bulk cannot be attributed
to the presence of an ordinary scalar field with a standard kinetic and
potential
term. An interpretation of such a bulk energy-momentum tensor was finally
given in terms of a scalar field with a non-linear kinetic term of the
form $(\partial_M \phi\,\partial^M \phi -c^2)^2$. By considering similar
models in the present case, as well as models with two scalar fields and
bulk Lagrangian of the form~\footnote{In all cases, the
potential $V$, either self-interacting or cross-interacting, had to
be trivially zero.}
${\cal L}_B \sim \partial_M \phi\,\partial^M \chi$, we were indeed able
to construct a bulk energy-momentum tensor with the desired non-vanishing
components. Although these models could also accommodate a $y$-dependent
black-hole horizon, none of the models constructed could support a decaying
horizon necessary for the localization of the black-hole singularity.
In principle, however, a model of scalar fields with a structure
$\partial_M\phi_a\,\partial^M\phi_b\,K_{ab} (\phi, \partial\phi)$
could lead to such a bulk energy-momentum tensor.

\section{Conclusions}

The construction of black-hole solutions in the context of
warped brane models has proven to be surprisingly difficult due to the
difficulty in localizing the black hole singularity near the brane. Although
some progress has been made in the numerical construction of such
solutions \cite{bh-nume},
a satisfactory analytical solution of this form, that would give us valuable
information regarding the nature of the gravitational collapse on a brane
embedded in a higher-dimensional spacetime, is still missing.

In Ref. \cite{KT} an attempt was made to localize the black-hole singularity
near the brane and, at the same time, avoid the singularity at the AdS horizon
by introducing
a dependence of the 4D horizon on the coordinate along the extra dimension.
An exponentially ``decaying" form of the black hole horizon in terms of
the $y$-coordinate was shown to achieve both of the aforementioned tasks.
Nevertheless, this non-factorizable line-element had to be sustained by a
bulk energy-momentum tensor, which although localized around the brane, was
of an exotic nature and was plagued by an additional singularity located at
the 4D horizon, a singularity that was also present in the expressions of the
curvature invariant quantities.

In this paper, we have pursued further the quest for localized black holes by
providing answers to questions that remained open after the end of our previous
work, such as: is the additional singularity a generic feature of every 
5D spacetime with a non-factorized line-element?can alternative line-elements be found, 
that are characterized by a similar scaling of the black hole horizon and
that do not possess this problem? does the corresponding bulk energy-momentum
tensor have to be always of an exotic nature?

We have started our analysis by demonstrating that non-factorized line-elements
which are not analytic at the horizon are always accompanied by the appearance of
singular terms in the expressions of the higher-dimensional curvature invariant
quantities. This leads to the manifestation of the 4D black-hole horizon as a
true spacetime singularity in the higher-dimensional spacetime. The above
result paved the way for the introduction on the brane of black-hole
line-elements related by coordinate transformations to the Schwarzschild one,
thus describing the same 4D spacetime, that were analytic
at the horizon. We considered two ans\"atze of this form, the Painlev\'e and
Vaidya line-elements, which were then embedded in a non-factorized way in a
5D warped brane model. In both cases, the resulting 5D spacetime was shown
to be well-defined for every $r>0$ while the scaling of the horizon with the
extra coordinate led to the localization of the black-hole singularity, at
$r=0$. In the case of the Painlev\'e line-element this scaling demands the
introduction of a regular, shell-type but highly complicated bulk energy-momentum
tensor that clearly cannot be attributed to a standard form of matter. On
the other hand, the Vaidya line-element proves to be a much more attractive,
natural choice. Despite the scaling of the horizon, the 5D spacetime has the
same topological structure as the one of a factorized RS-Schwarzschild spacetime.
This simplicity is also reflected in the form of the bulk energy-momentum
tensor: only two shell-like off-diagonal pressure components must
exist in the bulk while the energy-density and diagonal pressure components
are zero. Motivated by previously studied brane models with similar form of
bulk matter, we speculate that this particular bulk energy-momentum tensor may
be attributed not necessarily to an {\it exotic} form of matter but simply to
a {\it non-ordinary} one, i.e. scalar field theories with non-linear kinetic terms.

\bigskip

{\bf Acknowledgements.} 
I. O. and K. T. acknowledge the financial support of the EU RTN contract No.
HPRN-CT-2000-00152. K. T. acknowledges also the support of the EU RTN with
contract No. HPRN-CT-2000-00148. He also wishes to thank 
the CERN Theory Division for its hospitality.

\setcounter{equation}{0}
\def\theequation{A.\arabic{equation}}
\bigskip \bigskip
\noindent{\Large \bf Appendix}

\medskip

\noindent
For the $D$-dimensional spacetime described by the metric tensor given in
Eq. (\ref{metricD}), we obtain the following non-vanishing components of
the Christoffel symbols:
\bea
& \Gamma^\mu_{\rho\sigma} = \hat \Gamma^\mu_{\rho\sigma}\,, \qquad 
\Gamma^\mu_{\rho s}=\delta^\mu_\rho\,\partial_s A +
\frac{1}{2}\,\hat g^{\mu\nu} \hat g_{\rho\nu,s}\,, & 
\nonumber\\[4mm]
& \Gamma^m_{rs}= \tilde \Gamma^m_{rs}\,,
\qquad \Gamma^m_{\rho\sigma} = -\gamma^{mn}\,e^{2 A}\,(\partial_n A\,
\hat g_{\rho\sigma} + \frac{1}{2}\,\hat g_{\rho\sigma,n})\,,&
\eea
where $\hat \Gamma^\mu_{\rho\sigma}$ and $\tilde \Gamma^m_{rs}$ denote
the components of the Christoffel symbols being evaluated exclusively
in terms of the $\hat g_{\mu\nu}$ and $\gamma_{mn}$ metric tensor, respectively.
By using the above quantities, the relevant components of the $D$-dimensional
Ricci tensor are written as:
\bea
&~& \hspace*{-0.5cm}R_{\mu\nu} = \hat R_{\mu\nu} -\frac{e^{2A}}{2\sqrt{-g}}\,
\partial_m[\sqrt{-g}\,\gamma^{mn} \hat g_{\mu\nu,n}] 
+ \frac{1}{2}\,e^{2A}\,\hat g^{\rho\lambda}\,\gamma^{mn}\,
\hat g_{\mu\rho,m}\,\hat g_{\nu\lambda,n}
\\[2mm]
&~& \hspace*{0.3cm} 
-\,e^{2A}\,\hat g_{\mu\nu}\,\Bigl[(p+1)\,\partial^m A\,\partial_m A +
\tilde D^m\tilde D_m A + \partial^m A\,\partial_m \ln\sqrt{-\hat g}\Bigr]
-\frac{p+1}{2}\,e^{2A}\,\partial^m A\,\hat g_{\mu\nu,m} \nonumber 
%
\\[4mm]
&~& \hspace*{-0.5cm} 
R_{mn} = \tilde R_{mn} - \tilde D_m\tilde D_n \ln \sqrt{-\hat g} 
-\Bigl[\,\partial_m A\,\partial_n\ln\sqrt{-\hat g}
+\partial_n A\,\partial_m\ln\sqrt{-\hat g}\,\,\Bigr]
\nonumber \\[2mm]
&~& \hspace*{0.4cm}
-\,(p+1)\,\Bigl[\tilde D_m\tilde D_n A + \partial_m A\,\partial_n A\Bigr]
-\frac{1}{4}\,\hat g^{\alpha\beta}\,\hat g^{\rho\lambda}
\,\hat g_{\alpha\lambda,n}\,\hat g_{\rho\beta,m}\,,
\eea
where $\sqrt{-g}$ and $\sqrt{-\hat g}$ denote the determinant
of the $D$-dimensional and $(p+1)$-di\-men\-sio\-nal spacetime, respectively,
and $\tilde D_m$ the covariant derivative with respect to the
``extra" spacetime.

Contracting $R_{MN}$ by $g^{MN}$, we may determine the expression of the 
higher-di\-men\-sio\-nal Ricci scalar which is found to be
\bea
&~& \hspace*{-1.7cm} R = e^{-2A} \hat R + \tilde R - 
2 \tilde D^m \tilde D_m \ln\sqrt{-\hat g} -
\partial^m \ln\sqrt{-\hat g}\,\partial_m \ln \sqrt{-\hat g} +
\frac{1}{4}\,\hat g^{\mu\nu} \hat g^{\rho\lambda}\,\hat g_{\nu\lambda,m}\,
\hat g_{\mu\rho}^{\ \ \,,m}
\\[2mm] &~& \hspace*{-1.3cm}
+\,\,\frac{1}{2}\,\hat g_{\mu\nu,m}\,\hat g^{\mu\nu,m}
-(p+2)\,\partial^m A \Bigl[(p+1)\,\partial_m A+ 
2\partial_m \ln \sqrt{-\hat g}\,\Bigr] -2(p+1)\,\tilde D^m\tilde D_m A\,.
\nonumber  \label{scalarR}
\eea



\begin{thebibliography}{99}

\bibitem{large1}
I.~Antoniadis,
Phys.\ Lett.\ B {\bf 246}, 377 (1990);
N.~Arkani-Hamed, S.~Dimopoulos and G.~R.~Dvali,
Phys.\ Lett.\ B {\bf 429}, 263 (1998) [hep-ph/9803315];
Phys.\ Rev.\ D {\bf 59}, 086004 (1999) [hep-ph/9807344];
I.~Antoniadis, N.~Arkani-Hamed, S.~Dimopoulos and G.~R.~Dvali,
Phys.\ Lett.\ B {\bf 436}, 257 (1998)[hep-ph/9804398];


\bibitem{RS1}
L.~Randall and R.~Sundrum,
Phys.\ Rev.\ Lett.\  {\bf 83}, 3370 (1999)
[hep-ph/9905221].

\bibitem{RS2}
L.~Randall and R.~Sundrum,
Phys.\ Rev.\ Lett.\  {\bf 83}, 4690 (1999)
[hep-th/9906064].


\bibitem{graviton}
J.~Lykken and L.~J.~Randall,
JHEP {\bf 0006} (2000) 014 [hep-th/9908076];
K.~Skenderis and P.~K.~Townsend,
Phys.\ Lett.\ B {\bf 468} (1999) 46 [hep-th/9909070];
K.~Behrndt and M.~Cvetic, Phys.\ Lett.\ B {\bf 475} (2000) 253
[hep-th/9909058];
A.~Chamblin and G.~W.~Gibbons,
Phys.\ Rev.\ Lett.\  {\bf 84} (2000) 1090 [hep-th/9909130];
O.~DeWolfe, D.~Z.~Freedman, S.~S.~Gubser and A.~Karch,
Phys.\ Rev.\ D {\bf 62} (2000) 046008 [hep-th/9909134];
C.~Csaki, J.~Erlich, T.~J.~Hollowood and Y.~Shirman,
Nucl.\ Phys.\ B {\bf 581} (2000) 309 [hep-th/0001033];
S.~B.~Giddings, E.~Katz and L.~J.~Randall, JHEP {\bf 0003} (2000) 023
[hep-th/0002091];
H.~Kudoh and T.~Tanaka,
Phys.\ Rev.\ D {\bf 64} (2001) 084022 [hep-th/0104049];
Phys.\ Rev.\ D {\bf 67}, 044011 (2003) [hep-th/0205041].


\bibitem{bh-brane1}
R.~Emparan, G.~T.~Horowitz and R.~C.~Myers,
JHEP {\bf 0001} (2000) 007 [hep-th/9911043];
J.~Garriga and M.~Sasaki, Phys.\ Rev.\ D {\bf 62} (2000) 043523
[hep-th/9912118];
A.~Chamblin, C.~Csaki, J.~Erlich and T.~J.~Hollowood,
Phys.\ Rev.\ D {\bf 62} (2000) 044012 [hep-th/0002076];
N.~Dadhich, R.~Maartens, P.~Papadopoulos and V.~Rezania,
Phys.\ Lett.\ B {\bf 487} (2000) 1 [hep-th/0003061];
S.~Nojiri, O.~Obregon, S.~D.~Odintsov and S.~Ogushi,
Phys.\ Rev.\ D {\bf 62} (2000) 064017 [hep-th/0003148];
A.~Chamblin, H.~S.~Reall, H.~a.~Shinkai and T.~Shiromizu,
Phys.\ Rev.\ D {\bf 63} (2001) 064015 [hep-th/0008177];
I.~Giannakis and H.~c.~Ren,
Phys.\ Rev.\ D {\bf 63} (2001) 024001 [hep-th/0007053];
Phys.\ Rev.\ D {\bf 64} (2001) 065015 [hep-th/0103265];
Phys.\ Lett.\ B {\bf 528}, 133 (2002) [hep-th/0111127];
M.~S.~Modgil, S.~Panda and G.~Sengupta,
Mod.\ Phys.\ Lett.\ A {\bf 17}, 1479 (2002) [hep-th/0104122].

\bibitem{bh-brane2}
M.~Bruni, C.~Germani and R.~Maartens,
Phys.\ Rev.\ Lett.\  {\bf 87}, 231302 (2001) [gr-qc/0108013];
M.~Govender and N.~Dadhich,
Phys.\ Lett.\ B {\bf 538}, 233 (2002) [hep-th/0109086];
M.~Rogatko, Phys.\ Rev.\ D {\bf 64} (2001) 064014 [hep-th/0110018];
R.~Emparan and H.~S.~Reall,
Phys.\ Rev.\ Lett.\  {\bf 88}, 101101 (2002) [hep-th/0110260];
T.~Wiseman,
Phys.\ Rev.\ D {\bf 65}, 124007 (2002) [hep-th/0111057];
R.~Casadio, A.~Fabbri and L.~Mazzacurati,
Phys.\ Rev.\ D {\bf 65}, 084040 (2002) [gr-qc/0111072];
S.~I.~Vacaru and D.~Singleton,
Class.\ Quant.\ Grav.\  {\bf 19}, 3583 (2002) [hep-th/0112112];
R.~Casadio and L.~Mazzacurati,
Mod.\ Phys.\ Lett.\ A {\bf 18}, 651 (2003) [gr-qc/0205129];
R.~Emparan, A.~Fabbri and N.~Kaloper,
hep-th/0206155;
R.~Neves and C.~Vaz, 
Phys.\ Rev.\ D {\bf 66}, 124002 (2002) [hep-th/0207173];
P.~Kanti, I.~Olasagasti and K.~Tamvakis,
Phys.\ Rev.\ D {\bf 66}, 104026 (2002) [hep-th/0207283];
G.~Kofinas, E.~Papantonopoulos and V.~Zamarias,
Phys.\ Rev.\ D {\bf 66}, 104028 (2002) [hep-th/0208207];
R.~Neves and C.~Vaz,
Phys.\ Rev.\ D {\bf 68}, 024007 (2003) [hep-th/0302030];
S.~Shankaranarayanan and N.~Dadhich, gr-qc/0306111;
K.~A.~Bronnikov, V.~N.~Melnikov and H.~Dehnen, gr-qc/0304068;
R.~Neves and C.~Vaz,
Phys.\ Lett.\ B {\bf 568}, 153 (2003) [hep-th/0304266].


\bibitem{bh-nume}
T.~Wiseman,
Class.\ Quant.\ Grav.\  {\bf 20}, 1137 (2003) [hep-th/0209051];
H.~Kudoh, T.~Tanaka and T.~Nakamura, gr-qc/0301089.

\bibitem{large}
P.~C.~Argyres, S.~Dimopoulos and J.~March-Russell,
Phys.\ Lett.\ B {\bf 441}, 96 (1998) [hep-th/9808138];
S.~B.~Giddings and S.~Thomas,
Phys.\ Rev.\ D {\bf 65}, 056010 (2002) [hep-ph/0106219];
S.~Dimopoulos and G.~Landsberg,
Phys.\ Rev.\ Lett.\  {\bf 87}, 161602 (2001) [hep-ph/0106295];
R.~Emparan, G.~T.~Horowitz and R.~C.~Myers,
Phys.\ Rev.\ Lett.\  {\bf 85} (2000) 499 [hep-th/0003118];
M.~B.~Voloshin,
Phys.\ Lett.\ B {\bf 518}, 137 (2001) [hep-ph/0107119];
D.~M.~Eardley and S.~B.~Giddings,
Phys.\ Rev.\ D {\bf 66}, 044011 (2002) [gr-qc/0201034];
R.~Casadio and B.~Harms, 
Int.\ J.\ Mod.\ Phys.\ A {\bf 17}, 4635 (2002) [hep-th/0110255];
K.~Cheung, 
Phys.\ Rev.\ Lett.\  {\bf 88}, 221602 (2002) [hep-ph/0110163];
E.~J.~Ahn, M.~Cavaglia and A.~V.~Olinto,
Phys.\ Lett.\ B {\bf 551}, 1 (2003) [hep-th/0201042];
V.~Cardoso and J.~P.~Lemos,
Phys.\ Lett.\ B {\bf 538}, 1 (2002) [gr-qc/0202019]; 
A.~Chamblin and G.~C.~Nayak,
Phys.\ Rev.\ D {\bf 66}, 091901 (2002) [hep-ph/0206060];
V.~Frolov and D.~Stojkovic,
Phys.\ Rev.\ Lett.\  {\bf 89}, 151302 (2002) [hep-th/0208102];
P.~Kanti and J.~March-Russell,
Phys.\ Rev.\ D {\bf 66}, 024023 (2002) [hep-ph/0203223];
Phys.\ Rev.\ D {\bf 67}, 104019 (2003) [hep-ph/0212199];
Y.~Uehara, hep-ph/0205068; 
A.~V.~Kotwal and C.~Hays, Phys.\ Rev.\ D {\bf 66}, 116005 (2002)
[hep-ph/0206055];
T.~Han, G.~D.~Kribs and B.~McElrath,
Phys.\ Rev.\ Lett.\  {\bf 90}, 031601 (2003) [hep-ph/0207003];
D.~Ida, K.~y.~Oda and S.~C.~Park,
Phys.\ Rev.\ D {\bf 67}, 064025 (2003) [hep-th/0212108];
I.~Mocioiu, Y.~Nara and I.~Sarcevic,
Phys.\ Lett.\ B {\bf 557}, 87 (2003) [hep-ph/0301073];
A.~Chamblin, F.~Cooper and G.~C.~Nayak, hep-ph/0301239.


\bibitem{CHR}
A.~Chamblin, S.~W.~Hawking and H.~S.~Reall,
Phys.\ Rev.\ D {\bf 61}, 065007 (2000) [hep-th/9909205].


\bibitem{GL}
R.~Gregory and R.~Laflamme,
Phys.\ Rev.\ Lett.\  {\bf 70} (1993) 2837 [hep-th/9301052].

\bibitem{KT} P.~Kanti and K.~Tamvakis,
Phys.\ Rev.\ D {\bf 65}, 084010 (2002) [hep-th/0110298].



\bibitem{RS-AdS}
N.~Kaloper, Phys.\ Rev.\ D {\bf 60} (1999) 123506 [hep-th/9905210];
H.~B.~Kim and H.~D.~Kim,
Phys.\ Rev.\ D {\bf 61} (2000) 064003 [hep-th/9909053];
A.~Karch and L.~Randall,
JHEP {\bf 0105} (2001) 008 [hep-th/0011156];
T.~Hirayama and G.~Kang,
Phys.\ Rev.\ D {\bf 64} (2001) 064010 [hep-th/0104213].



\bibitem{PAIN} P. Painlev{\'{e}}, C. R. Acad. Sci. (Paris), {\bf{173}},
677(1921).

\bibitem{KW} P. Kraus and F. Wilczek, Mod. Phys. Lett. {\bf{A 9}}, 3713
(1994);
M.~K.~Parikh and F.~Wilczek,
Phys.\ Rev.\ Lett.\  {\bf 85}, 5042 (2000) [hep-th/9907001].

\bibitem{vaidya}
P.C. Vaidya, Nature {\bf 171} (1953) 260. 

\bibitem{PI} E.~Poisson and W.~Israel,
Phys.\ Rev.\ D {\bf 41} (1990) 1796;
C.~Barrabes and W.~Israel,
Phys.\ Rev.\ D {\bf 43} (1991) 1129.

\bibitem{KKOP}
P.~Kanti, I.~I.~Kogan, K.~A.~Olive and M.~Pospelov,
Phys.\ Lett.\ B {\bf 468} (1999) 31 [hep-ph/9909481];
Phys.\ Rev.\ D {\bf 61} (2000) 106004 [hep-ph/9912266].

\bibitem{KOP} C.~Csaki, M.~Graesser, L.~Randall and J.~Terning,
Phys.\ Rev.\ D {\bf 62}, 045015 (2000) [hep-ph/9911406];
P.~Kanti, K.~A.~Olive and M.~Pospelov,
Phys.\ Lett.\ B {\bf 481}, 386 (2000) [hep-ph/0002229];
Phys.\ Rev.\ D {\bf 62}, 126004 (2000) [hep-ph/0005146].


\end{thebibliography}
\end{document}